\documentstyle[preprint,aps]{revtex}
\begin{document}

\draft
\title{ A note on radiative corrections to $\mu$ and $\tau$ decays }

\author{ V. Konyshev }
\address{ Landau Institute for Theoretical Physics, 119740, Moscow, Russia }

\author{ I. Polyubin }
\address{ Landau Institute for Theoretical Physics, 119740, Moscow,%
Russia\\ and \\
ITEP, 117259, Moscow, Russia }

\date{\today}
\maketitle

\begin{abstract}
Radiative corrections in the order
${\alpha\over{2\pi}}{{m^2_e}\over m^2_\mu}$ to $\mu $- and
${\alpha\over{2\pi}}{{m^2_\mu}\over m^2_\tau}$ to $\tau $- decays
are calculated. The decay width is enhanced by
$4.48\cdot 10^{-3} ({\alpha\over{2\pi}})$ in the muon case and by
$0.283 ({\alpha\over{2\pi}})$ for the $\tau \rightarrow \mu \nu_\tau %
\bar\nu_\mu (\gamma)$
decay. Influence of these corrections on the electroweak data is discussed.
\end{abstract}


The nowaday level of accuracy in the electroweak experiments requires the
complete accounting of the $O(\alpha)$ radiative corrections \cite{1}.
Any deviation of the
experimental data from these predictions can be considered as an indication
of the presence of some new physics beyond the Standard Model. Therefore,
the careful determination of these corrections is of great importance.

In this letter we reconsider the one-loop electromagnetic corrections to the
muon and $\tau$ decays \cite{2}. We calculate the contributions in the order
${\alpha\over{2\pi}}{{m^2_e}\over m^2_\mu}$ to the total $\mu$ decay rate and
${\alpha\over{2\pi}}{{m^2_\mu}\over m^2_\tau}$
to the muon decays of $\tau$. The order of magnitude of this
correction has been discussed in \cite{3} (see also \cite{4}) and has been
estimated to be of order $10^{-7}\div 10^{-8}$.
We integrate the well-known one-loop radiatively corrected electroweak
spectrum \cite{2} in muon decay numerically and extract the first term
of the expansion in ${{m^2_e}\over{m^2_\mu}}$.
It turns out that the numerical factor in front of
this term enhances this correction by two orders. With accounting this
correction the total decay rate of muon is equal to
\begin{equation}\label{c}
\Gamma(\mu\rightarrow\text{all}) = {{G^2_F m^5_\mu}\over{192\pi^3}}%
\left[{ f\!\left({{m^2_e}\over{m^2_\mu}}\right) + %
{3\over 5}{{m^2_\mu}\over{m^2_W}} + %
{{\alpha({m_\mu})}\over{2\pi}}g\!\left({{m^2_e}\over{m^2_\mu}}\right) %
}\right] ,
\end{equation}
\begin{equation}
f(x) = 1 - 8x + 8x^3 - x^4 - 12x^2 \log x ,
\end{equation}
\begin{equation}\label{d}
g(x) = {25\over 4} - {\pi}^2 - 24x \left(\log x +{17\over 6}\right) + o(x) ,
\end{equation}
\begin{equation}
\alpha^{-1}({m_\mu}) = \alpha^{-1} - {1\over{3\pi}} %
\log {{m^2_\mu}\over{m^2_e}} + {1\over{6\pi}} \approx 136.1
\end{equation}
with
\begin{equation}
m_\mu = 105.658389 \pm 0.000034 \text{ MeV },
\end{equation}
\begin{equation}
m_e = 0.51099906 \pm 0.00000015 \text{ MeV },
\end{equation}
\begin{equation}
m_W = 80.22 \pm 0.26 \text{ GeV }.
\end{equation}
Thus, the corrections are:
\begin{mathletters}
\begin{equation}
{3\over 5} {{m^2_\mu}\over{m^2_W}} \approx 1.04 \cdot 10^{-6} ,
\end{equation}
\begin{equation}
\left(g\!\left({{m^2_e}\over{m^2_\mu}}\right) - g(0)\right) %
{\alpha\over{2\pi}} \approx 4.48 \cdot 10^{-3} %
{\alpha\over{2\pi}} \approx 5.2 \cdot 10^{-6} ,
\end{equation}
\begin{equation}\label{a}
{{\alpha^2}\over{3\pi}} \log {{m^2_\mu}\over{m^2_e}} \left( %
\pi^2 - {25\over 4} \right) \approx 3.5 \cdot 10^{-5} .
\end{equation}
\end{mathletters}

One can see that this correction is five times larger than the correction
due
to the finite $W$ mass. Nevertheless, this correction is irrelevant for the
concrete value of the weak constant
$G_F = (1.16639\pm 0.00002)\cdot 10^{-5} \text{GeV}^{-2}$
\cite{1}, since it is 7 times smaller than (\ref{a}) which is quoted as
the theoretical uncertainty for the weak constant \cite{1}.

Indeed, the decay rate for $\Gamma( \tau \rightarrow e \nu_\tau \bar\nu_e %
(\gamma) )$ and for  $\Gamma( \tau \rightarrow \mu \nu_\tau \bar\nu_\mu %
(\gamma) )$ can be simply obtained from (\ref{c}) by the replacement
$m_\mu \rightarrow m_\tau$ and $\left( m_e \rightarrow m_\mu \text{; } %
m_\mu \rightarrow m_\tau \right)$ respectively. The correction under
consideration is relevant for $\Gamma(\tau \rightarrow \mu \nu_\tau %
\bar\nu_\mu (\gamma))$. Explicitly, one finds
\begin{equation}
\Gamma( \tau \rightarrow \mu \nu_\tau \bar\nu_\mu (\gamma) ) = %
{{G^2_F m^5_\tau}\over{192\pi^3}}%
\left[{ f\!\left({{m^2_\mu}\over{m^2_\tau}}\right) + %
{3\over 5}{{m^2_\tau}\over{m^2_W}} + %
{{\alpha({m_\tau})}\over{2\pi}}g\!\left({{m^2_\mu}\over{m^2_\tau}}\right) %
}\right]
\end{equation}
with
\begin{equation}
\alpha^{-1}(m_\tau) \approx 133.3 ,
\end{equation}
\begin{equation}
m_\tau = 1777.0 \pm 0.26 \text{ MeV (\cite{5}) .}
\end{equation}
Then, one gets
\begin{mathletters}
\begin{equation}
{3\over 5} {{m^2_\tau}\over{m^2_W}} \approx 2.89 \cdot 10^{-4} ,
\end{equation}
\begin{equation}
\left(g\!\left({{m^2_\mu}\over{m^2_\tau}}\right) - g(0)\right) %
{\alpha\over{2\pi}} \approx 0.283 \cdot %
{\alpha\over{2\pi}} \approx 3.25 \cdot 10^{-4} ,
\end{equation}
\begin{equation}
{{\alpha^2}\over{3\pi}} \log {{m^2_\tau}\over{m^2_e}} \left( %
\pi^2 - {25\over 4} \right) \approx 5.3 \cdot 10^{-5} ,
\end{equation}
\end{mathletters}
Let us note that the first term in the expansion of $g({x})$ in (\ref{d})
gives $g({{m^2_\mu}\over{m^2_\tau}}) - g(0) \approx 0.23 $.

In the case of $\tau \rightarrow \mu \nu_\tau \bar\nu_\mu (\gamma)$
decay the account of nonzero muon mass decreases the
one-loop correction by 8\%.
This correction slightly affects on predictions
for branching ratios of the $\tau$ leptonic decays:
\begin{equation}\label{b}
\Gamma(\tau \rightarrow \mu \nu_\tau \bar\nu_\mu (\gamma) ) = 0.9731 \cdot%
\Gamma(\tau \rightarrow e \nu_\tau \bar\nu_e (\gamma) ).
\end{equation}

This correction also relevant for testing of $e-\mu$ universality in $\tau$
leptonic decays. Rewriting (\ref{b}) in the form \cite{5}:
\begin{equation}
\Gamma(\tau \rightarrow \nu_\tau l \bar\nu_l (\gamma) ) = %
{{G_\tau G_l m^5_\tau}\over{192\pi^3}}\left[ f({x}_l) + %
{3\over 5} {{m^2_\tau}\over{m^2_W}} + %
{{\alpha(m_\tau)}\over{2\pi}}g({x}_l) \right] ,
\end{equation}
where $G_l ={{g^2_l}\over{4\sqrt{2}M^2_W}}, {x}_l = %
{{m^2_l}\over{m^2_\tau}}$.

The strength of each leptonic charged current is determined by $g_l$.
The comparison of $\Gamma_e$ and $\Gamma_\mu$ is a test of $e-\mu$
universality. Since $f({x}_e) \approx 1$ , $g({x}_e) \approx 0 $, one has:
\begin{equation}
{{\Gamma_\mu}\over{\Gamma_e}} = f({x}_\mu) {{g^2_\mu}\over{g^2_e}}
\end{equation}
with $f({x}_\mu) \approx 0.9731$.

Comparing with the experimental data \cite{5}, one finally gets
\begin{equation}
{{g_\mu}\over{g_e}} = 1.0005 \pm 0.0035
\end{equation}

To conclude, we compute correction of the order
$({\alpha\over{2\pi}}){{m^2_e}\over{m^2_\mu}}$ to the muon width and of the
order $({\alpha\over{2\pi}}){{m^2_\mu}\over{m^2_\tau}}$ to
$\Gamma (\tau \rightarrow \mu \nu_\tau \bar\nu_\mu (\gamma ) )$ respectively.
In the last case, this correction decreases one-loop correction by 8\%.
Our results give the precise predictions for the decay rates, which can be
really tested provided the experiments achieve higher accuracy and the
explicit calculation of the two-loop correction for the total muon decay rate
is done.

I.~Polyubin thanks L.B.~Okun for useful discussions. This work was supported
in part by RFFI grant N~93-011-16087, RFFI grant N~94-02-14365 and
ISF grant MET000.

{\bf Note Added}. After this work was completed we have learned that the
formula (\ref{d}) already exists in literature \cite{6}. We are grateful to
Y.Nir for bringing the paper \cite{6} to our attention.

\end{document}